\documentclass[pra,twocolumn,showkeys,showpacs,nofootinbib,endfloats*]{revtex4-1}

\usepackage{amsmath,amssymb,amsfonts,amsbsy,bm}   
\usepackage{graphicx,graphics,color,epsfig,subfig}
\usepackage{lastpage,fancyhdr}                    
\usepackage{float}                                
\usepackage{dcolumn}                              
\usepackage[sort&compress]{natbib}                
\usepackage{color}                                
\usepackage[normalem] {ulem}

\DeclareGraphicsExtensions{.eps,.tif}    

\newcommand{\half}{\frac{1}{2}}

\newcommand{\bra}{\langle}
\newcommand{\ket}{\rangle}

\begin{document}
\title{Symmetric Rotating Wave Approximation for the Generalized Single-Mode Spin-Boson System}
\author{Victor V. Albert}
\affiliation{Department of Chemistry and Centre for Quantum Information and Quantum Control, University of Toronto, Toronto, Canada M5S 3H6}
\author{Gregory D. Scholes}
\affiliation{Department of Chemistry and Centre for Quantum Information and Quantum Control, University of Toronto, Toronto, Canada M5S 3H6}
\author{Paul Brumer}
\email[Corresponding author: ]{pbrumer@chem.utoronto.ca}
\affiliation{Department of Chemistry and Centre for Quantum Information and Quantum Control, University of Toronto, Toronto, Canada M5S 3H6}
\date{\today}
\begin{abstract}

The single-mode spin-boson model exhibits behavior not
included in the rotating wave approximation (RWA) in the
ultra  and  deep-strong  coupling regimes, where
counter-rotating contributions become important. We
introduce a symmetric rotating wave approximation that
treats rotating and counter-rotating terms equally,
preserves the invariances of the Hamiltonian with respect to
its parameters, and reproduces several qualitative features
of the spin-boson spectrum not present in the original
rotating wave approximation both off-resonance and at deep
strong coupling. The symmetric rotating wave approximation
allows for the treatment of certain ultra and deep-strong
coupling regimes with similar accuracy and mathematical
simplicity as does the RWA in the weak coupling regime.
Additionally, we symmetrize the generalized form of the
rotating wave approximation to obtain the same qualitative
correspondence with the addition of improved quantitative
agreement with the exact numerical results. The method is
readily extended to higher accuracy if needed. Finally, we
introduce the two-photon parity operator for the two-photon
Rabi  Hamiltonian  and  obtain its generalized symmetric
rotating wave approximation. The existence of this operator
reveals a parity symmetry similar to that in the Rabi
Hamiltonian as well as another symmetry that is unique to
the two-photon case, providing insight into the mathematical
structure  of  the  two-photon  spectrum,  significantly
simplifying the numerics, and revealing some interesting
dynamical properties.



\end{abstract}
\maketitle

The Hamiltonian of a two-level system coupled linearly to a quantum harmonic oscillator \cite{fg_original, jc} is one of the most studied models in quantum mechanics and is still unsolved despite over fifty years of nearly continuous effort. The model has found applications in many fields ranging from molecular chemistry to circuit quantum electrodynamics, which is evident when one lists the different pseudonyms for it in the literature: the Jaynes-Cummings model (without the rotating wave approximation) \cite{dimer_single_mode_chen_recursion_arxiv, Shore1993, *dimer_one_mode_larson, *Zhang2011, *dimer_one_mode_naderi} and the single-mode spin-boson \cite{dimer_one_mode_feng, math2, dimer_integrability, *Hagelstein2008} in condensed matter physics, the Rabi Hamiltonian in quantum optics \cite{dimer_one_mode_Bishop_crosspts, *spin_boson_another_unitary, *dimer_werlang, *Travenec2011}, the molecular dimer-oscillator in chemical physics \cite{wagner, dimer_engel}, and the spin-oscillator \cite{spin_boson_trwa} and qubit-oscillator in quantum information and circuit QED \cite{spin_boson_ultrastrong, spin_boson_bloch_siegert, dimer_one_mode_obada, *grifoni1}. Many analytical approaches have been developed for small and (more recently) large coupling between the system and the oscillator, with arguably the most well-known being the rotating wave approximation (RWA). In this article we introduce and discuss approximations using similar techniques as those used in obtaining the RWA, but in parameter regions complementing it.

The RWA \cite{jc} is designed to work well in the case of weak coupling between the two-level system and the oscillator. Additionally, previous approximations \cite{dimer_one_mode_talab_earlier, dimer_single_mode_Feranchuk_earliest} have been linked to a generalized form of the RWA in Ref. \cite{dimer_one_mode_irish_grwa_prl}. However, recent experiments \cite{Ciuti2005, *grifoni15, *Schoelkopf2008, *Bourassa2009, *Wilson2010, *Alton2010} have motivated the study of this model in parameter regions which have not been thoroughly explored in the past and have shown that the RWA breaks down in those regions \cite{JC_nature, spin_boson_bloch_siegert}. Specifically, a number of recent theoretical studies have shown that contributions of counter-rotating terms, which are ignored in the RWA, prove important in these parameter regions and exhibit dynamical behavior different from the weak-coupling case \cite{spin_boson_counter_rotating, spin_boson_deep_strong_coupling, grifoni2, spin_boson_ultrastrong, jing2009, *zueco2009, *spin_boson_variational, *spin_boson_switchable, *Zhang2011a, *Beaudoin2011}. Additionally, counter-rotating terms dominate the short-time dynamical behavior for some parameter regions, leading to important Zeno and anti-Zeno effects that are not reproduced by the RWA \cite{spin_boson_zeno, spin_boson_zeno_physics, Ai2010, *Ai2010a, *Cao2010}.

The Hamiltonian of a two-level system coupled \textit{quadratically} to a quantum harmonic oscillator, the two-photon Rabi Hamiltonian, has also been studied within the RWA \cite{tprh_rwa_2, *tprh_old}. Limitations of the RWA have been outlined for this system \cite{tprh_rwa_foundation}, but so far limited effort has been directed to studying it outside of the RWA \cite{tprh_bishop_exact, emary_thesis_tprh_dimer, Peng1993, tprh_special_functions, *tprh_algebras_dolya}. This Hamiltonian arose in quantum optics as a phenomenological model for a three-level system interacting with two photons \cite{tprh_rwa_foundation, emary_thesis_tprh_dimer, tprh_buck} and is also relevant in modeling pure dephasing in crystals \cite{Skinner1986, *Skinner1988}. As opposed to a displacement in position in the case of the Rabi Hamiltonian, the coupling in the two-photon Rabi Hamiltonian is through frequency displacement or ``squeezing'' \cite{perelomov_book, tprh_diagonalization}. With bi-exciton effects and two-photon processes occurring in experimental systems \cite{Brune1987, *two_photon_in_QDs, *tprh_application, *DelValle2010, *Ota2011}, more work is needed to determine whether this Hamiltonian can successfully model these effects.

The two Hamiltonians ($m=1,2$) can be written in the form
\begin{equation}\label{eq:1}
H_{m}=\omega b^{\dagger}b+J\sigma_{x}+\lambda\sigma_{z}\left(b^{m}+\left(b^{\dagger}\right)^{m}\right)
\end{equation}
where $H_1$ is unitarily equivalent to the single-mode spin boson/Rabi Hamiltonian and $H_2$ is the two-photon Rabi Hamiltonian. Here, $\sigma_i$ are the usual Pauli matrices \cite{dimer_one_moder_small_j_irish}, $b^\dagger$ and $b$ are the boson raising and lowering operators, $\omega$ is the harmonic oscillator frequency, $J$ is the coupling of the two-level system, and $\lambda$ is the coupling strength between the two-level system and the harmonic oscillator. The recent regions of interest for $H_1$ include the ultra-strong ($\lambda \gtrsim 0.1 \omega$) and deep-strong ($\lambda \gtrsim \omega$) coupling regimes \cite{spin_boson_deep_strong_coupling, spin_boson_ultrastrong}.

In this work, we introduce a symmetric form of the rotating wave approximation (denoted as S-RWA) that includes an equal amount of rotating and counter-rotating terms. The S-RWA provides analogous physical insight in the off-resonance ultra and deep-strong coupling regions to that provided by the RWA at resonance in the weak coupling limit. Similar to previous use of the RWA in the weak-coupling limit \cite{Narozhny1981, *scully}, one can employ the S-RWA to extract important features of the dynamics in the strong coupling limit. In Section \ref{sec1}, we discuss a mathematical feature of $H_1$ conserved by the S-RWA: the invariance of its energies under change of sign of the coupling parameters $J$ and $\lambda$. In Section \ref{sec2}, we symmetrize the RWA and describe the parameter regions in which it is most applicable. In Section \ref{sec3}, we provide a generalization of the S-RWA. Similar to the generalization of the RWA to larger coupling \cite{dimer_one_mode_irish_grwa_prl}, the analogous generalization of the S-RWA (denoted as S-GRWA) extends it to regions of small coupling (sketched in Table \ref{t1}). The S-GRWA can also be extended to higher levels of accuracy if needed, a feature that is not as directly evident in the generalized RWA. Finally, by introducing the two-photon parity operator in Section \ref{sec4}, we are able to apply analogous methods and obtain an S-GRWA for the two-photon Rabi Hamiltonian $H_2$. We show that, apart from respecting the same symmetries present in $H_1$, the two-photon case contains two independent manifolds and the two-photon parity operator maintains parity symmetries on each of them.

\begin{table}
\begin{center}
\begin{tabular}{cccc}
  \hline
  \hspace*{10pt}\\[-8pt]
  & $~~\lambda \ll \omega~~$ & & $~~\lambda \gtrsim \omega~~$ \\[2pt]
  \hline
  \hspace*{10pt}\\[-8pt]
  $\omega \approx 2J$   &  RWA & $\rightarrow$ & GRWA \\
  $~~\omega \gg 2J~~$   & S-GRWA & $\leftarrow$ & S-RWA \\
  \hline
\end{tabular}\end{center}
\caption{Sketch of the RWA and S-RWA and their respective extensions in the parameter space. The arrows represent the direction of extension upon generalization of each respective approximation.\label{t1}}
\end{table}

\section{Symmetries in Bosonic and Fermionic Systems}\label{sec1}

Changing the basis for a Hamiltonian $H$ is equivalent to applying a unitary transformation $U$ to obtain a transformed Hamiltonian $\widetilde{H}=U^{\dagger}HU$, with the eigenvalues of $H$ preserved \cite{wagner_book}. Therefore, if there exists a unitary transformation such that applying that transformation corresponds to changing the sign of a parameter of $H$, then the set of eigenvalues must be invariant under this change of sign.

We note two relevant examples of this property. The first is the degenerate two-level system: $H=\epsilon+J\sigma_{x}$. Applying the two-level fermionic reflection operator $U_{r}\left(\pi\right)$ \cite{wagner_book}, where
\begin{equation}
U_{r}\left(\phi\right)=\exp\left(i\phi\sigma_{y}/2\right),
\end{equation}
is equivalent to letting $J \rightarrow -J$. As expected, the set of eigenvalues of this Hamiltonian, $\left\{\epsilon \pm J\right\}$, is invariant under this transformation. The analogous boson problem is the displaced harmonic oscillator
\begin{equation}\label{eq:d}
H_1^+=\omega b^{\dagger}b+\lambda\left(b+b^{\dagger}\right),
\end{equation}
which can be diagonalized with the unitary displacement operator \cite{perelomov_book}
\begin{equation}\label{eq:disp}
\mathcal{D}\left(\lambda/\omega\right)=\exp\left[\left(b-b^{\dagger}\right) \lambda/\omega \right],
\end{equation}
yielding the set of energies $\{\omega N-\frac{\lambda^{2}}{\omega}\}_{N=0}^{\infty}$. Applying the well-known bosonic parity/reflection operator,
\begin{equation}
\mathcal{R}=\exp\left(i\pi b^{\dagger}b\right),
\end{equation}
is equivalent to letting $\lambda \rightarrow -\lambda$. In this case, the invariance is more pronounced: while the two-level system involves a re-ordering of the index $\left\{\epsilon \pm J\right\} \rightarrow \left\{\epsilon \mp J\right\}$, the energies of both the original and transformed boson Hamiltonians are equal for each $N$ due to their dependence on $\lambda^2$. 

For $H_m$ from Eq. (\ref{eq:1}), both of these invariances are present and can be confirmed numerically. For the Rabi Hamiltonian $H_1$, applying the transformations $\mathcal{R}$ and $U_r(\pi) \mathcal{R}$ is equivalent to letting $\lambda \rightarrow -\lambda$ and $J \rightarrow -J$, respectively. It is important to note that the invariance with respect to $J$ is not a mere convention: for any real $J$, symmetrically correct sets of energy approximations have to be the same for both $J$ and $-J$. The exact energies of $H_1$ could therefore be dependent on even powers of $J$ and/or contain a ``$\pm$'' splitting for odd powers. While these concepts may seem trivial, the invariance with respect to $J$, while maintained for some newer approximations \cite{dimer_single_mode_chen_recursion_arxiv, grifoni2, dimer_one_moder_small_j_irish}, is broken for the RWA \cite{jc}, GRWA \cite{dimer_one_mode_irish_grwa_prl}, and other approximations \cite{dimer_one_mode_feng, spin_boson_trwa, dimer_single_mode_Feranchuk_earliest, feranchuk2}, as discussed below.

\section{Symmetrizing the Rotating Wave Approximation}\label{sec2}

The RWA consists of rotating the system by $U_r(\pi/2)$ and then transforming into the interaction picture with respect to the interaction-free Hamiltonian $H^{(0)}=\omega b^{\dagger}b + J \sigma_x$. Assuming $J > 0$, the terms which rotate at frequencies of $\omega + 2J$ are assumed to oscillate much faster than the terms rotating at $\omega - 2J$ and are thus ignored. Note that the counter-rotating wave approximation (CRWA) ignores the slower-oscillating terms, i.e., $\omega - 2J$, and obtains the excited state energies of the RWA with the sign of $J$ changed \cite{dimer_isolated_solutions}. The RWA energies are $E_{0}^{\text{RWA}}=-J$ and (for $N>0$)
\begin{equation}
E_{N,\pm}^{\text{RWA}}=\left(N+\half\right)\omega\pm\sqrt{\left(\half\omega-J\right)^{2}+\lambda^{2}\left(N+1\right)}.
\end{equation}
Note that the RWA energies are proportional to a difference between $\frac{1}{2}\omega$ and $J$, known in quantum optics as the detuning frequency. Thus, any given set of RWA (or CRWA) energies is not invariant under changing the sign of $J$ due to the symmetry breaking that these approximations cause. An additional effect of this is the existence of an isolated ground state that does not depend on the coupling $\lambda$, which has been pointed out previously \cite{spin_boson_parity_breaking, dimer_one_mode_feng}. Using the methods outlined below, we obtain the symmetric RWA energies (with $p=\pm$ and for $N \geq 0$)
\begin{equation}\label{eq:swa}
E_{N,p,\pm}^{\text{S-RWA}}=\left(2N+\half\right)\omega\pm\sqrt{\left(\half\omega-pJ\right)^{2}+\lambda^{2}\left(2N+1\right)}.
\end{equation}
The extra ``parity index'' $p$ is precisely the reason for invariance under the sign of $J$. The isolated ground state disappears and the S-RWA ground state, $E^{\text{S-RWA}}_{0,-,-}$, is of the same accuracy as the excited states. It should also be noted that the excited RWA (CRWA) states for even $N$ are exactly the positive (negative) parity S-RWA states, demonstrating that the symmetrization maintains the relative mathematical simplicity of the RWA while also including rotating and counter-rotating terms equally.

\subsection{Obtaining the S-RWA}

\begin{figure}[htpp]
 \begin{center}
 \includegraphics[width=0.45\textwidth]{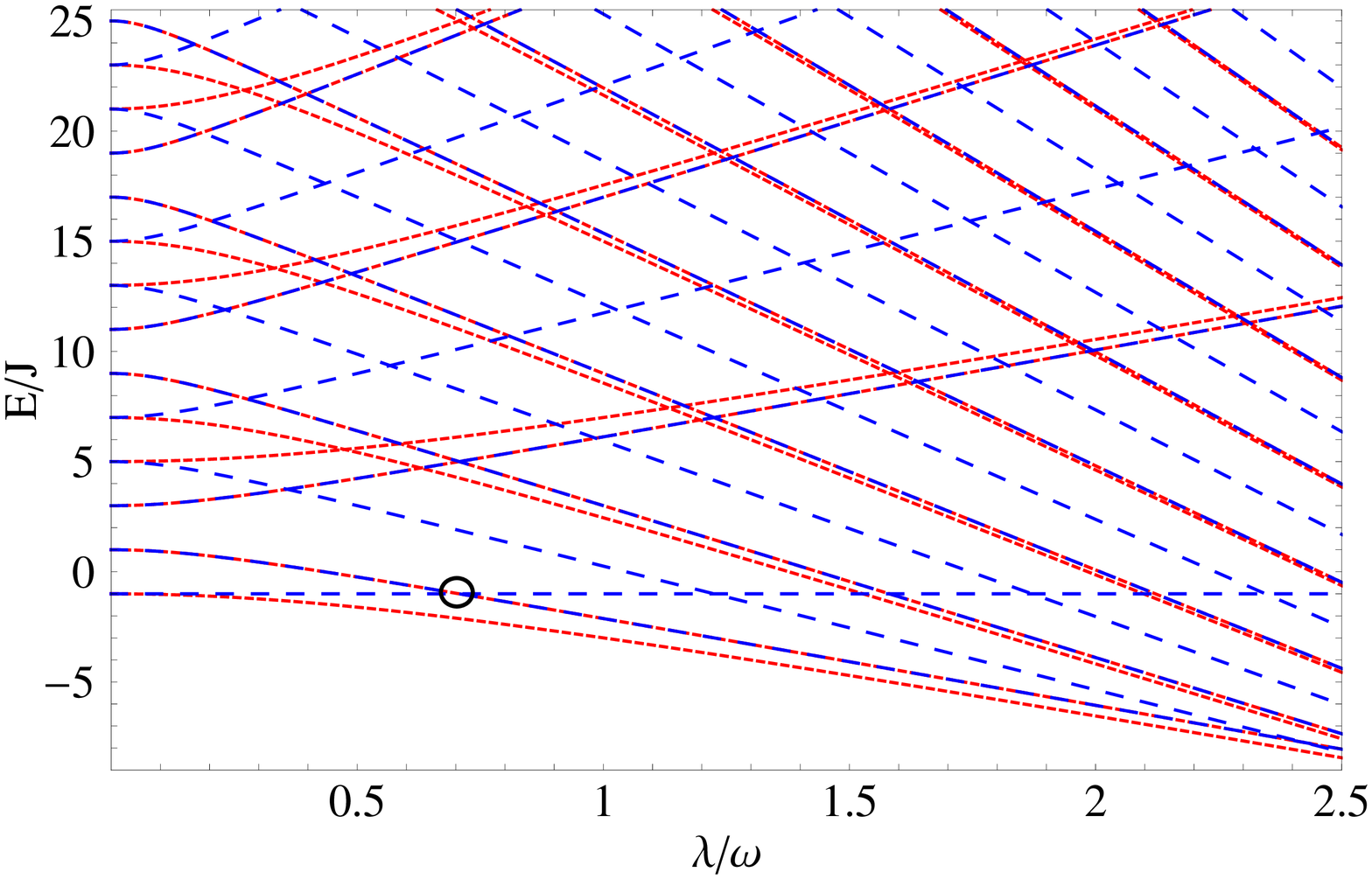}
 \end{center}
 \caption{Comparison of the first few RWA (long blue dashes) and S-RWA (short red dashes) energies at $4\omega = J$. The two features of the exact spectrum present in the S-RWA and not in the RWA are convergence of energies of opposing parity at large coupling and the ability of the S-RWA to properly retain the ground state for large values of $\lambda$. The black circle depicts where the RWA ground state is surpassed by an excited state. \label{f1}}
\end{figure}

To obtain Eq. (\ref{eq:swa}), note that Eq. (\ref{eq:1}) can be written explicitly in the spin-$\frac{1}{2}$ basis $|\pm\ket$ (with $|+\ket\bra+|$ being the top left entry), where each of the four entries are operators on the boson Fock space. In this form, both Hamiltonians are
\begin{equation}\label{eq:fg}
H_{m}=\left(\begin{array}{cc}
H_{m}^{+} & J\\
J & H_{m}^{-}\end{array}\right)
\end{equation}
where $H_{m}^{\pm}=\omega b^{\dagger}b\pm\lambda\left(b^{m}+\left(b^{\dagger}\right)^{m}\right)$ are symmetric matrices in the Fock (number-state) basis. Instead of rotating them by $U_r(\pi/2)$ as prescribed by the RWA, we introduce a non-local unitary transformation
\begin{equation}\label{eq:un}
U_{m}=\frac{1}{\sqrt{2}}\left(\begin{array}{cc}
1 & -\mathcal{P}_{m}\\
\mathcal{P}_{m}^{\dagger} & 1\end{array}\right)
\end{equation}
that will diagonalize these Hamiltonians in the spin subspace, with the unitary $\mathcal{P}_m$ determined below.
In order for the two off-diagonal terms of $\widehat{H}_m=U_m^{\dagger} H_m U_m$ to vanish, the following two equations must be satisfied:
\begin{eqnarray}
\mathcal{P}_{m}^{\dagger}H_{m}^{+}\mathcal{P}_{m}&=&H_{m}^{-} \label{eq:cond} \\
\mathcal{P}_{m}^{2}&=&1. \label{eq:cond2}
\end{eqnarray}
The second equation shows that $\mathcal{P}_m$ is in fact a parity operator: $\mathcal{P}_{m}=\mathcal{P}_{m}^{-1}=\mathcal{P}_{m}^{\dagger}$. If such a $\mathcal{P}_m$ exists, then the transformed Hamiltonian $\widehat{H}_m$ is diagonal in the spin subspace:
\begin{equation}
\widehat{H}_{m}=\left(\begin{array}{cc}
H_{m}^{+}+J\mathcal{P}_{m} & 0\\
0 & H_{m}^{-}-J\mathcal{P}_{m}\end{array}\right).
\end{equation}
Various forms of this transformation for $H_1$ have been proposed independently \cite{fg_original, wagner, spin_boson_parity_breaking, spin_boson_deep_strong_coupling, spin_boson_shore_sander_fg, *wagner_fg} and the transformation is extendable to the multi-mode case. For $m=1$, Ref. \cite{fg_original} gives $\mathcal{P}_1 = \mathcal{R}=\exp\left(i\pi b^{\dagger}b\right)$ and the Hamiltonian $\widehat{H}_1$ consists of the diagonal entries
\begin{equation}\label{eq:fg_m}
\widehat{H}_1^{\pm}=\omega b^{\dagger}b\pm\lambda\left(b+b^{\dagger}\right)\pm J\mathcal{R}
\end{equation}
which are both Hamiltonian operators on the boson Fock space. The $\widehat{H}_1^{\pm}$ provide analytical and numerical advantages over the original form of the problem \cite{paganelli2, dimer_one_mode_pan, wagner, crwa_monster_hohnerbach, spin_boson_deep_strong_coupling, math2}. Additionally, they differ only by the signs of the parameters and correspond to parity-related subspaces \cite{wagner} or ``parity chains'' \cite{spin_boson_deep_strong_coupling}. Writing $\widehat{H}_1^{\pm}$ in the Fock state basis $|N\ket$ (where $N=0,1,2,...$), we make the approximation by truncating them into two sets of 2-by-2 diagonal blocks
\[
\left(\begin{array}{cc}
2N\omega\pm J & \pm\lambda\sqrt{2N+1}\\
\pm\lambda\sqrt{2N+1} & \left(2N+1\right)\omega\mp J\end{array}\right).
\]
The eigenvalues of these blocks are the symmetric RWA energies given in Eq. (\ref{eq:swa}), a result that we call the S-RWA. The S-RWA is plotted with the RWA in Fig. \ref{f1} at $J=\frac{1}{4}\omega$. At a given $\lambda$ in the ultra-strong and (even more so) deep strong coupling regimes and in the off-resonance cases ($J < \frac{1}{2}\omega $), energies of opposing parity approach one another with increasing $N$ and the exact numerical spectrum of the spin-boson can be divided into two nearly degenerate columns of differing parity. In these regimes, and in this limit, parity no longer determines the energy but merely defines the column to which a given energy belongs. It is thus applicable and appropriate to analyze each parity case separately at deep strong coupling \cite{spin_boson_deep_strong_coupling}. The failure of the RWA at ultra-strong coupling can be seen in Fig. \ref{f1}. In the limit of large $\lambda$, the respective upward and downward-sloping RWA energies are equally spaced while the S-RWA energies of opposing parity pair up and correctly approach one another ($E_{N,p,\pm}^{\text{S-RWA}}\approx E_{N,-p,\pm}^{\text{S-RWA}}$). Additionally, the RWA ground state is surpassed by the first excited state at $\lambda \approx 0.7$ (black circle), resulting in an un-physical flipping of the well-defined ground state parity not present in the exact results. On the other hand, the S-RWA ground state is maintained throughout the entire parameter range shown. The trade-off to the improved behavior at large coupling is the presence of un-physical crossings of the negative parity S-RWA excited state energies at $\lambda < 0.5$. In order to better understand the origin of these features and the physical relation of the S-RWA to the RWA, we examine the two approximations analytically in the next section.

\subsection{Comparison with the RWA}

When performing the RWA, $H_1$ is written in the eigenbasis of $\sigma_x$ coupled to a Fock state:
\begin{equation}
|\pm x,N\ket=\frac{1}{\sqrt{2}}\left(|+\ket\pm|-\ket\right)\otimes|N\ket.
\end{equation}
There are only two types of off-diagonal terms in this basis: resonant $ \bra+x,N|H_{1}|-x,N+1\ket $ and off-resonant $ \bra-x,N|H_{1}|+x,N+1\ket $ (and their respective conjugate transposes). The RWA (CRWA) ignores the off-resonant (resonant) contributions and keeps the resonant (non-resonant) terms \cite{dimer_one_mode_irish_grwa_prl}.
In the case of the S-RWA, the transformation $U_1$ allows a different partitioning of these off-diagonal terms, making it possible to symmetrically ignore half of the resonant and half of the off-resonant contributions. Rotating by $U_1$ is equivalent to writing $H_1$ in the basis
\begin{equation}
|\phi_{\pm,N}\ket=\frac{1}{\sqrt{2}}\left(|\pm,N\ket\pm\left(-1\right)^{N}|\mp,N\ket\right).
\end{equation}
The original basis $|\pm x,N\ket$ is thus mixed in a such a way that $\bra\phi_{\pm,N} |H_1| \phi_{\mp,M}\ket=0$ for all $N,M$. We can therefore write $H_1$ separately in terms of $|\phi_{+,N}\ket$ and $|\phi_{-,N}\ket$, resulting in $\widehat{H}_1^+$ and $\widehat{H}_1^-$, respectively. Similar to the RWA, we now observe that the off-diagonal terms can be partitioned into two types: $\bra\phi_{\pm,2N}| H_{1} |\phi_{\pm,2N+1}\ket$ and $\bra\phi_{\pm,2N+1}| H_{1} |\phi_{\pm,2N+2}\ket$. The first type consists of even-to-odd Fock-space transitions which couple $ |\pm x,2N\ket $ and $ |\mp x,2N+1\ket $ while the second are odd-to-even transitions coupling $ |\pm x,2N+1\ket $ and $ |\mp x,2N+2\ket $ (up to a $\pm$ overall phase). The S-RWA keeps the first type and ignores the second, which is equivalent to removing half of the resonant and half of the off-resonant terms.


The RWA is an approximation that is valid for the resonance case ($\omega\approx2J$) and at small coupling since rotating terms dominate the long-time dynamical behavior in this region: $\omega-2J\ll\omega+2J$. One the other hand, the S-RWA is the analogous approximation in the complementary off-resonance ($\omega \gg 2J$) and deep-strong coupling region, where $\omega-2J \approx \omega+2J$. In this regime, both the rotating and counter-rotating terms are equally as important \cite{spin_boson_zeno}. This leads to novel short-time dynamics and the Zeno effect in the single \cite{spin_boson_zeno_physics} and multi-mode spin-boson systems \cite{spin_boson_zeno, Ai2010, *Ai2010a, *Cao2010}, topics of significant interest. While neither the RWA nor the S-RWA energies from Eq. (\ref{eq:swa}) are particularly accurate in quantitatively reproducing the full numerical spectrum, generalizing the S-RWA, in a method analogous to generalizing the RWA, produces improved agreement with the exact results at resonance and at large coupling.

\section{Symmetrizing the Generalized RWA}\label{sec3}

The generalization of the RWA involves a simple change of
basis involving $\mathcal{D}$ from Eq. (\ref{eq:disp}) prior
to performing the analogous 2-by-2 matrix truncation
performed in the RWA. The GRWA thus includes behaviors of
the   adiabatic   approximation
\cite{dimer_one_mode_irish_grwa_prl}, extending the validity
of the RWA to arbitrarily large couplings. Since the S-RWA
is valid for only large values of the coupling, the
symmetric generalized RWA extends the validity of the S-RWA
to arbitrarily small coupling in the off-resonance regime.
The regions of applicability of the respective extensions of
the RWA and S-RWA are qualitatively depicted in
Table \ref{t1}.

\begin{figure}[htpp]
 \begin{center}
 \subfloat(a){\includegraphics[width=0.45\textwidth]{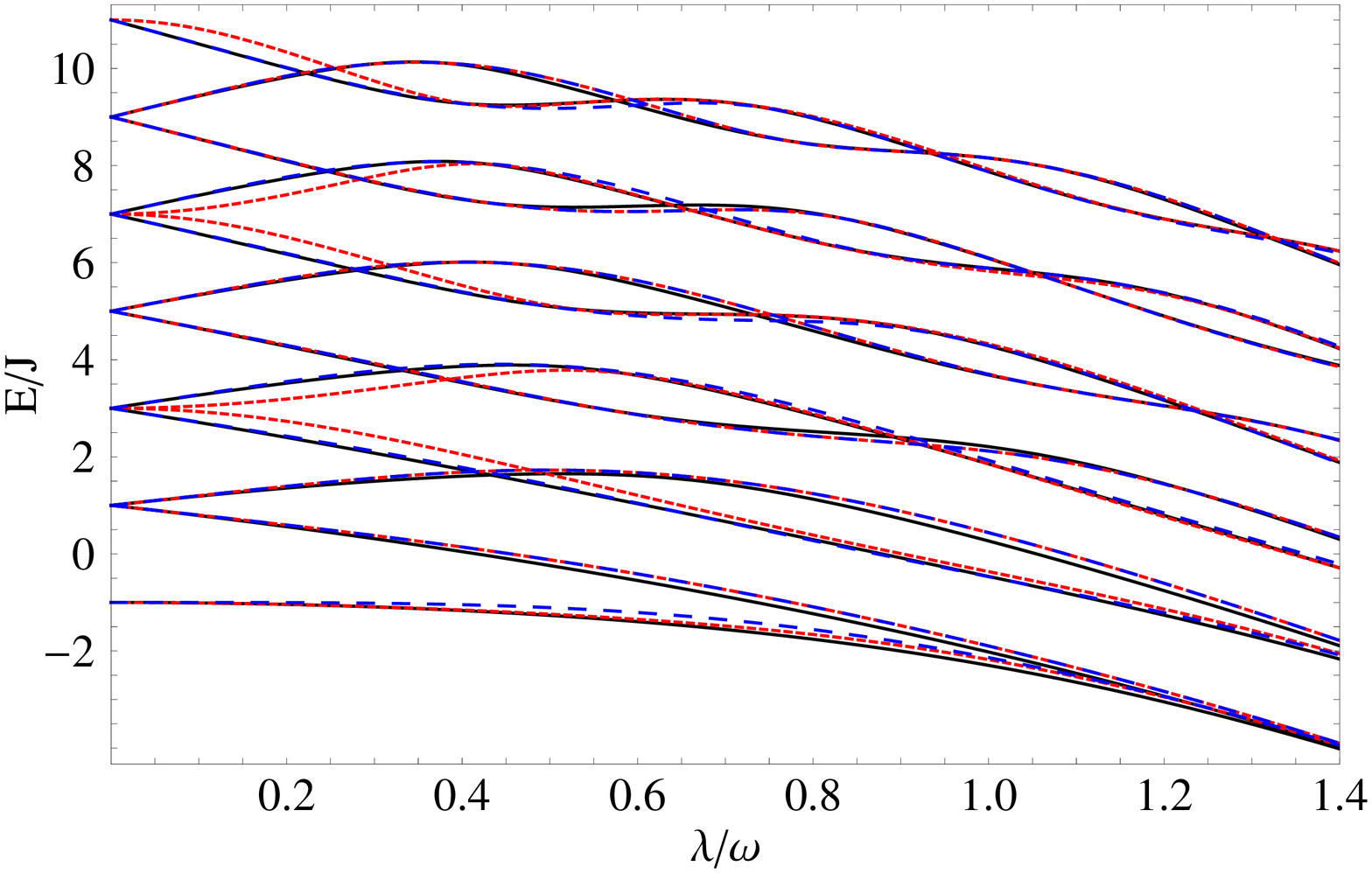}}\\ 
 \subfloat(b){\includegraphics[width=0.45\textwidth]{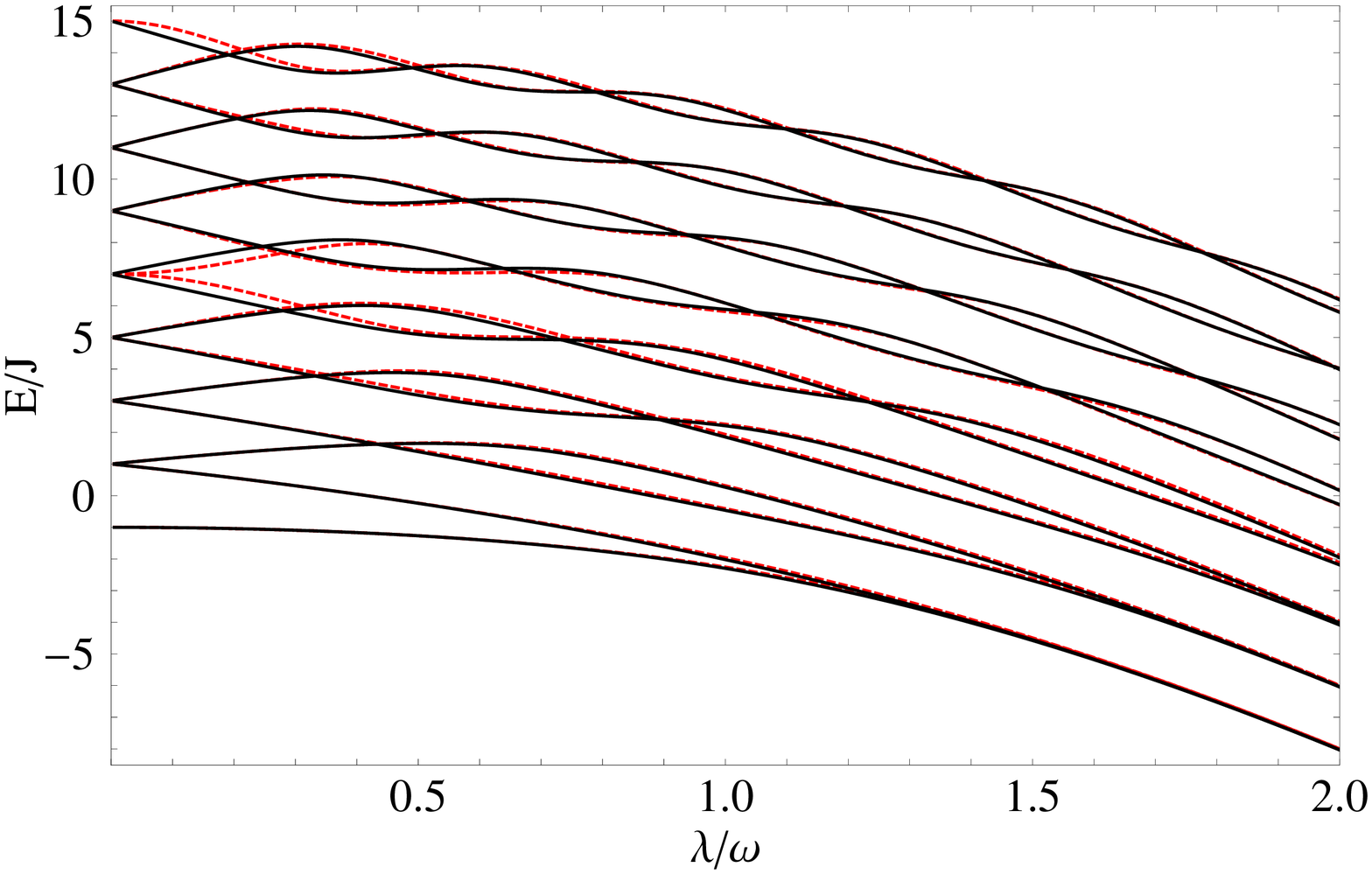}}
 \end{center}
 \caption{(a) Numerical energies (solid black) compared with the 2-by-2 S-GRWA (short red dashes) and GRWA (long blue dashes) at the resonance case $ 2J = \omega = 1 $, where the differences are most evident. (b) The 4-by-4 S-GRWA matrix truncation (dashed) plotted against the numerical energies (solid). The initial physical reasoning for the respective RWA and S-RWA approximations suggest that the GRWA should be used for the small coupling regime while the S-GRWA, with its respect for parity and a corrected ground state, should be used for $\lambda > 0.5$. \label{f2}}
\end{figure}

The derivation of the S-GRWA is similar to that of the GRWA since a similar change of basis is performed in both. The diagonal $\widehat{H}_1$ is now transformed with $U_{\mathcal{D}}=\mathcal{D}\left(\frac{\lambda}{\omega}\sigma_{z}\right)$, removing the linear bosonic coupling terms in $\widehat{H}^{\pm}_1$, obtaining \cite{paganelli2}
\begin{equation}
\widetilde{H}_{1}^{\pm}=\omega b^{\dagger}b -\lambda^{2}/\omega \pm J\mathcal{RD}\left(\pm2\lambda/\omega\right).
\end{equation}
These Hamiltonians are written solely in terms of the number operator and two forms of the displaced parity operator \cite{spin_boson_parity}, which differ only in the displacement direction. To write them in Fock space, we need the Fock space matrix elements of the displacement operator \cite{perelomov_book}, $D_{M,N} = \bra M| \mathcal{D}(2\lambda/\omega) |N\ket$, which are\footnote{Here, $L_{M}^{N-M}(x)$ is an associated Laguerre polynomial, which can be defined for all values of integers $M$ and $N$ \cite{perelomov_book}. Since $\mathcal{D}(-2\lambda/\omega)=\mathcal{D}^\dagger(2\lambda/\omega)$, the elements $\bra M|\mathcal{D}(-2\lambda/\omega)|N\ket = D_{N,M}$.}
\begin{equation}
D_{M,N}=\sqrt{\frac{M!}{N!}}e^{-2\lambda^{2}/\omega^{2}} \hspace{-1mm} \left(\frac{2\lambda}{\omega}
\right)^{N-M} \hspace{-1mm} L_{M}^{N-M}\left(\frac{4\lambda^{2}}{\omega^{2}}\right).
\end{equation}
To obtain the symmetric GRWA, we write $\widetilde{H}_{1}^{\pm}$ in Fock space and, mimicking the method utilized in the generalized RWA \cite{dimer_one_mode_irish_grwa_prl}, truncate them to 2-by-2 block-diagonal form. Excluding the ``$-\frac{\lambda^{2}}{\omega}$'' term, the $N$th block is then given by
\begin{equation}
\left(\begin{array}{cc}
2N\omega\pm JD_{2N,2N} & JD_{2N,2N+1}\\
JD_{2N,2N+1} & \left(2N+1\right)\omega\mp JD_{2N+1,2N+1}\end{array}\right)\label{eq:sgrwa}.
\end{equation}
Since the transformed matrix is full (as opposed to having many zero entries as is the case of the GRWA), there is the additional liberty of choosing the size of the matrix truncation. Surprisingly, energies identical to the 1-by-1 block diagonal truncation have been obtained previously via other methods \cite{crwa_monster_hohnerbach, dimer_one_mode_feng, dimer_one_mode_talab_earlier}. We also note that the 2-by-2 and 4-by-4 truncations were performed in a very similar treatment \cite{dimer_one_mode_feng}, but the negative parity excited state energies were discarded. The 2-by-2 truncation performed above is of the same order of accuracy as the GRWA, and diagonalizing Eq. (\ref{eq:sgrwa}) gives the following S-GRWA energies:
\begin{widetext}
\begin{eqnarray}
E_{N,p,\pm}^{\text{S-GRWA}}&=&\left(2N+\half\right)\omega-\frac{\lambda^{2}}{\omega}+p\frac{J}{2}e^{-2\lambda^{2}/\omega^{2}}\left[L_{2N}\left(4\lambda^{2}/\omega^{2}\right)-L_{2N+1}\left(4\lambda^{2}/\omega^{2}\right)\right]\\&\pm&\left(\left\{ \half\omega-p\frac{J}{2}e^{-\frac{2\lambda^{2}}{\omega^{2}}}\left[L_{2N}\left(4\lambda^{2}/\omega^{2}\right)+L_{2N+1}\left(4\lambda^{2}/\omega^{2}\right)\right]\right\} ^{2}+\frac{4\lambda^{2}J^{2}}{\omega^{2}\left(2N+1\right)}e^{-4\lambda^{2}/\omega^{2}}\left[L_{2N}^{1}\left(4\lambda^{2}/\omega^{2}\right)\right]^{2}\right)^{1/2}
\nonumber
\end{eqnarray}
\end{widetext}
As with the positive parity ($p=+$) S-RWA energies being equal to the even RWA excited state energies, the positive parity S-GRWA energies are exactly the even GRWA excited state energies. The numerical (solid black) and approximated energies (dashed) are plotted against the coupling $\lambda$ in Fig. \ref{f2}(a) for the resonance case $2J = \omega = 1$, where differences between them are most evident. At couplings $\lambda > 0.5$, the agreement is almost identical with the GRWA, with the exception of the improved ground state of the S-GRWA. The applicability of the S-RWA is vastly extended to the regions of smaller coupling $\lambda < 0.5$, but the GRWA still maintains better agreement with the exact results in that region. The negative parity energies, alternating by two with the positive parity energies, are not in good agreement with the exact results at small coupling. In order to correct this, one can easily program the 4-by-4 matrix truncation into \textsc{Mathematica}, which has been done in Fig. \ref{f2}(b) for a larger coupling range. Here, half of the artifacts of the negative parity disappear, occurring in two out of every eight excited state energies as opposed to every four. In both truncations, these artifacts decrease with increasing energy level and disappear in the off-resonance regime, making the 4-by-4 truncation virtually indistinguishable from the exact energies in that regime (not shown). The initial physical reasoning for the respective RWA and S-RWA approximations suggest that the GRWA should be used for the small coupling regime while the S-GRWA, with its respect for parity and a corrected ground state, should be used for $\lambda > 0.5$.

\section{Two-Photon Rabi Hamiltonian}\label{sec4}

\subsection{Even-odd separation}

Consider the two-photon case described by $H_2$ from Eq. (\ref{eq:1}). We first note a symmetry of this system that is not present in $H_1$, namely that $H_2$ does not couple even and odd Fock states. This has been discussed in a different context \cite{tprh_special_functions, *tprh_algebras_dolya} and stems from the nature of the quadratic coupling in the squeezed harmonic oscillator $H_2^\pm$ from Eq. (\ref{eq:fg}). Writing $H_2^\pm$ in the Fock state basis, we arrive at a pentagonal matrix with two middle diagonals equal to zero:
\begin{equation}\nonumber
\left(\begin{array}{ccccc}
\fbox{0} & 0 & \fbox{$\pm\lambda\sqrt{2}$} & 0 & \\
0 & \dashbox{3}(15,15){$\omega$} & 0 & \dashbox{3}(38,20){$\pm\lambda\sqrt{6}$} & \\
\fbox{$\pm\lambda\sqrt{2}$} & 0 & \fbox{$2\omega$} & 0 & \ddots\\
0 & \text{\dashbox{3}(38,20){$\pm\lambda\sqrt{6}$}} & 0 & \dashbox{3}(21,15){$3\omega$} & \ddots\\
 & & \ddots & \ddots & \ddots\end{array}\right).
\end{equation}
The square (dashed square) boxes represent the even (odd) manifold couplings; all other couplings are zero. This is equivalent to stating that $ \bra 2N | H_2^\pm | 2M+1 \ket $ and $ \bra 2N+1 | H_2^\pm | 2M \ket$ are zero for all $N,M$. Thus, one can reorder the basis and write $H_2^\pm$ first in the even basis and then in the odd, converting the pentagonal into two tridiagonal matrices: 
\begin{equation}\nonumber
\left(\begin{array}{cccccc}
0 & \pm\lambda\sqrt{2} & 0\\
\pm\lambda\sqrt{2} & 2\omega & \ddots & & \textbf{0}\\
0 & \ddots & \ddots\\
 & & & \omega & \pm\lambda\sqrt{6} & 0 \\
 & \textbf{0} & & \pm\lambda\sqrt{6} & 3\omega & \ddots\\
 & & & 0 & \ddots & \ddots\end{array}\right).
\end{equation}
It is important to stress that the even and odd manifolds are orthogonal and independent of each-other. This property admits an interesting observation: if the initial wavefunction for this system contains only even (odd) Fock states, then it's evolved wavefunction will remain in the even (odd) Fock state manifold. Additionally, if the initial wavefunction is comprised of both even and odd Fock states, then those states will evolve separately and will not exchange populations with one another at any time.

This property of $H_2$ is conserved under unitary transformation, is independent of the values of the parameters, and will be used to obtain symmetric GRWA energies for this system. As a result, there will be one set of energies for each manifold, as opposed to just one set for $H_1$. In addition to this even-odd decoupling, the invariances with respect to $J$ and $\lambda$ that hold for $H_1$ also hold for $H_2$, as shown below.

\subsection{Two-photon parity}

We introduce the two-photon parity operator:
\begin{equation}
\mathcal{T}=\exp\left[i\pi \frac{1}{2} b^{\dagger}b \left(b^{\dagger}b-1\right)\right].
\end{equation}
Acting on a Fock state $|N\ket$, the operator can be written as follows:
\begin{equation}
\mathcal{T}|N\ket=\left(-1\right)^{\half N\left(N-1\right)}|N\ket=\exp\left(i\pi\sum_{j=0}^{N-1}j\right)|N\ket.
\end{equation}
Note that the eigenvalue of $\mathcal{T}$ for a state $|N\ket$ is a product of the parities of the previous $N-1$ states. Since $\frac{1}{2}N(N-1)$ is always an integer, it is easy to prove that $ \mathcal{T} = \mathcal{T}^\dagger = \mathcal{T}^{-1} $. It is evident that $\mathcal{T}$ maintains the same behavior for the manifold of even and odd Fock states as $\mathcal{R}$ does for the set of all states:
\begin{equation}
\mathcal{T}=\sum_{N=0}^{\infty}\left(-1\right)^{N}\left(|2N\ket\bra2N|+|2N+1\ket\bra2N+1|\right).
\end{equation}
This is not accidental, given that $H_2$ is decoupled within the even and odd state subspaces as mentioned above. Thus, there is a parity on each of the even/odd subspaces, giving a total of four parities: $\{ -1_e,-1_o,+1_e,+1_o \}$.\footnote{We note that the Fourier operator $\sqrt{\mathcal{R}}$ has been labeled as the ``pseudo-parity'' of $H_2$ \cite{emary_thesis_tprh_dimer, tprh_bishop_exact, spin_boson_parity} because the true parity has previously been unknown.}


The two-photon parity $\mathcal{T}$ commutes with $\mathcal{R}$ and its actions on the raising and lowering operators are:
\begin{eqnarray}
\mathcal{T}b&=&-b\mathcal{R}\mathcal{T}\\
\mathcal{T}b^{\dagger}&=&b^{\dagger}\mathcal{R}\mathcal{T}.\nonumber
\end{eqnarray}
We know from Eq. (\ref{eq:cond}) that $ \mathcal{R}b\mathcal{R} = -b$ and $ \mathcal{R}b^\dagger\mathcal{R} = -b^\dagger$. Thus, $ \mathcal{T}$ satisfies Eq. (\ref{eq:cond}) and anti-commutes with the quadratic raising and lowering operators:
\begin{eqnarray}
\mathcal{T}b^2&=&-b^2\mathcal{T}\\
\mathcal{T}b^{\dagger 2}&=&-b^{\dagger 2}\mathcal{T}\nonumber.
\end{eqnarray}
The two-photon parity $\mathcal{T}$ thus satisfies both conditions imposed by Eqs. (\ref{eq:cond}) and (\ref{eq:cond2}), implying that it is indeed the parity operator $\mathcal{P}_2$ needed to diagonalize the Hamiltonian in the spin subspace using the transformation $U_2$ from Eq. (\ref{eq:un}). Similar to $H_1$, the two-photon Rabi Hamiltonian energies are also invariant under change in sign of $\lambda$ and $J$. Applying the transformations $\mathcal{T}$ and $U_r(\pi) \mathcal{T}$ is equivalent to letting $\lambda \rightarrow -\lambda$ and $J \rightarrow -J$, respectively. In analogy with the displaced harmonic oscillator $H_1^+$ from Eq. (\ref{eq:d}), the squeezed harmonic oscillator $H_2^+$ is diagonalized by the unitary squeeze operator \cite{wagner_book}
\begin{equation}
\mathcal{S}\left(\alpha\right)=\exp\left[\textstyle\frac{1}{2}\alpha\left(b^{2}-b^{\dagger2}\right)\right],
\end{equation}
producing the set of energies $ \left\{ \tilde{\omega}N-\half\left(\omega-\tilde{\omega}\right)\right\} _{N=0}^{\infty} $, where
\begin{equation}
\tilde{\omega}\equiv\omega\sqrt{1-\left(2\lambda/\omega\right)^{2}}.\label{eq:upb}
\end{equation}
The parameter $\alpha$ in $\mathcal{S}(\alpha)$ satisfies the conditions
\begin{eqnarray}\label{eq.cond}
\sinh 2\alpha &=&2\lambda/\tilde{\omega}\\\cosh2\alpha &=&\omega/\tilde{\omega}.\nonumber
\end{eqnarray}
Equation (\ref{eq:upb}) implies an upper bound for the squeezed coupling: $\lambda < \frac{1}{2}\omega $, and thus $ 0 <\tilde{\omega}\leq\omega $. More importantly, the energies of the squeezed harmonic oscillator are invariant under change of sign of $\lambda$, as is the case of the displaced oscillator. Thus, $H_2$ preserves all of the invariances of $H_1$ and contains the additional ability to be divided into even and odd Fock state manifolds.

\subsection{S-GRWA for the two-photon case}

We perform the same steps as with $H_1$ in Section \ref{sec3}, except that now the two-photon parity is used ($\mathcal{R} \rightarrow \mathcal{T}$) and squeezed as opposed to displaced Fock states are implemented ($\mathcal{D} \rightarrow \mathcal{S}$). First, one obtains $\widehat{H}_2 = U_2^\dagger H_2 U_2$ with diagonal entries
\begin{equation}\label{eq:fg2}
\widehat{H}_{2}^{\pm}=\omega b^{\dagger}b\pm\lambda\left(b^{2}+b^{\dagger2}\right)\pm J\mathcal{T}.
\end{equation}
The Hamiltonian $H_2$ is thus divided into two parity-related subspaces. Furthermore, each of the parity spaces can be divided into even and odd subspaces, giving one subspace per each of the four parities of $\mathcal{T}$. The four operators are tridiagonal matrices in Fock space and significantly decrease the effort required to obtain accurate numerical energies for $H_2$, paving the way for the efficient application of diagonalization schemes \cite{dimer_one_mode_pan}. Additionally, this four-fold separation reveals that diagonalizing $H_2$ is equivalent to determining the spectrum of an infinite tridiagonal matrix with all three diagonals approaching infinity, as is the case for $H_1$ \cite{math2}. The differences are that the diagonals increase faster and there are four as opposed to two matrices due to the extra even/odd manifold symmetry. Finally, the Hamiltonian $H_1$ commutes with the parity operator $\mathcal{R}\sigma_x$, a combination of the inversion operators for the respective spin a displaced boson systems. The Hamiltonian $H_2$ commutes with both $\mathcal{T}\sigma_x$ and $\mathcal{R}$. The reflection operator determines the manifold ($e$ or $o$) while $\mathcal{T}\sigma_x$ determines the parity on each manifold ($\pm1$), giving the four parities mentioned above.



Utilizing 2-by-2 block diagonals of the four tridiagonal matrices in Eq. (\ref{eq:fg2}) gives the symmetric RWA for $H_2$. In order to obtain the more accurate generalized version of these energies, we apply $U_{\mathcal{S}}=\mathcal{S}\left(\alpha\sigma_{z}\right)$ to further transform $\widehat{H}_{2}^{\pm}$ into:
\begin{equation}\label{eq:fg3}
\widetilde{H}_{2}^{\pm}=\tilde{\omega}b^{\dagger}b -\textstyle\half\left(\omega-\tilde{\omega}\right) \pm J \mathcal{T}\mathcal{S}\left( \pm2\alpha \right).
\end{equation}
Here, we use the fact that $ \mathcal{T}\mathcal{S}\mathcal{T}=\mathcal{S}^{\dagger} $, just like $ \mathcal{RDR}=\mathcal{D}^{\dagger} $ in the linear case. The squeeze operator is also separable into even and odd subspaces, and its Fock space matrix elements $S_{M,N} = \bra M|\mathcal{S}(2\alpha)|N\ket$  for real $\alpha > 0$ are:\footnote{Here, $P_{M+N}^{M-N}(x)$ is an associated Legendre polynomial, which can be defined for all values of integers $M$ and $N$. Since $\mathcal{S}(-\alpha)=\mathcal{S}^\dagger(\alpha)$, the elements $\bra M|\mathcal{S}(-\alpha)|N\ket = S_{N,M}$. These were obtained using Ref. \cite{silbey_1978} and Eq. (\ref{eq.cond}); they are in agreement with previous efforts \cite{perel, *Satyanarayana, *P.Mari}.}
\begin{eqnarray}
S_{2M,2N}&\hspace{-1mm}=&\sqrt{\frac{\left(2N\right)!}{\left(2M\right)!}}\left(\frac{\widetilde{\omega}}{\omega}\right)^{\half}
  \hspace{-1mm} P_{M+N}^{M-N}\hspace{-1mm}\left(\frac{\widetilde{\omega}}{\omega}\right)\\
S_{2M+1,2N+1}&\hspace{-1mm}=&\sqrt{\frac{\left(2N+1\right)!}{\left(2M+1\right)!}}\left(\frac{\widetilde{\omega}}{\omega}\right)^{\half}
  \hspace{-1mm}P_{M+N+1}^{M-N}\hspace{-1mm}\left(\frac{\widetilde{\omega}}{\omega}\right)\hspace{-1mm}.
\end{eqnarray}
Writing $\widetilde{H}_{2}^{\pm}$ in the Fock basis, dividing into the two manifolds, and truncating to 2-by-2 diagonal form (although higher truncations are again possible) gives the following four matrices (excluding the ``$-\half\left(\omega-\tilde{\omega}\right)$'' term):
\small
\begin{eqnarray}
\nonumber&
\left(\begin{array}{cc}
4N\tilde{\omega}\pm J S_{4N,4N} & J S_{4N,4N+2}\\
J S_{4N,4N+2} & \left(4N+2\right)\tilde{\omega}\mp J S_{4N+2,4N+2}\end{array}\right) &\\
\nonumber&
\left(\begin{array}{cc}
\left(4N+1\right)\tilde{\omega}\pm J S_{4N+1,4N+1} & J S_{4N+1,4N+3}\\
J S_{4N+1,4N+3} & \left(4N+3\right)\tilde{\omega}\mp J S_{4N+3,4N+3}\end{array}\right).
\end{eqnarray}
\vspace{1mm}
\normalsize

\begin{figure}[htpp]
 \begin{center}
 \subfloat(a){\includegraphics[width=0.45\textwidth]{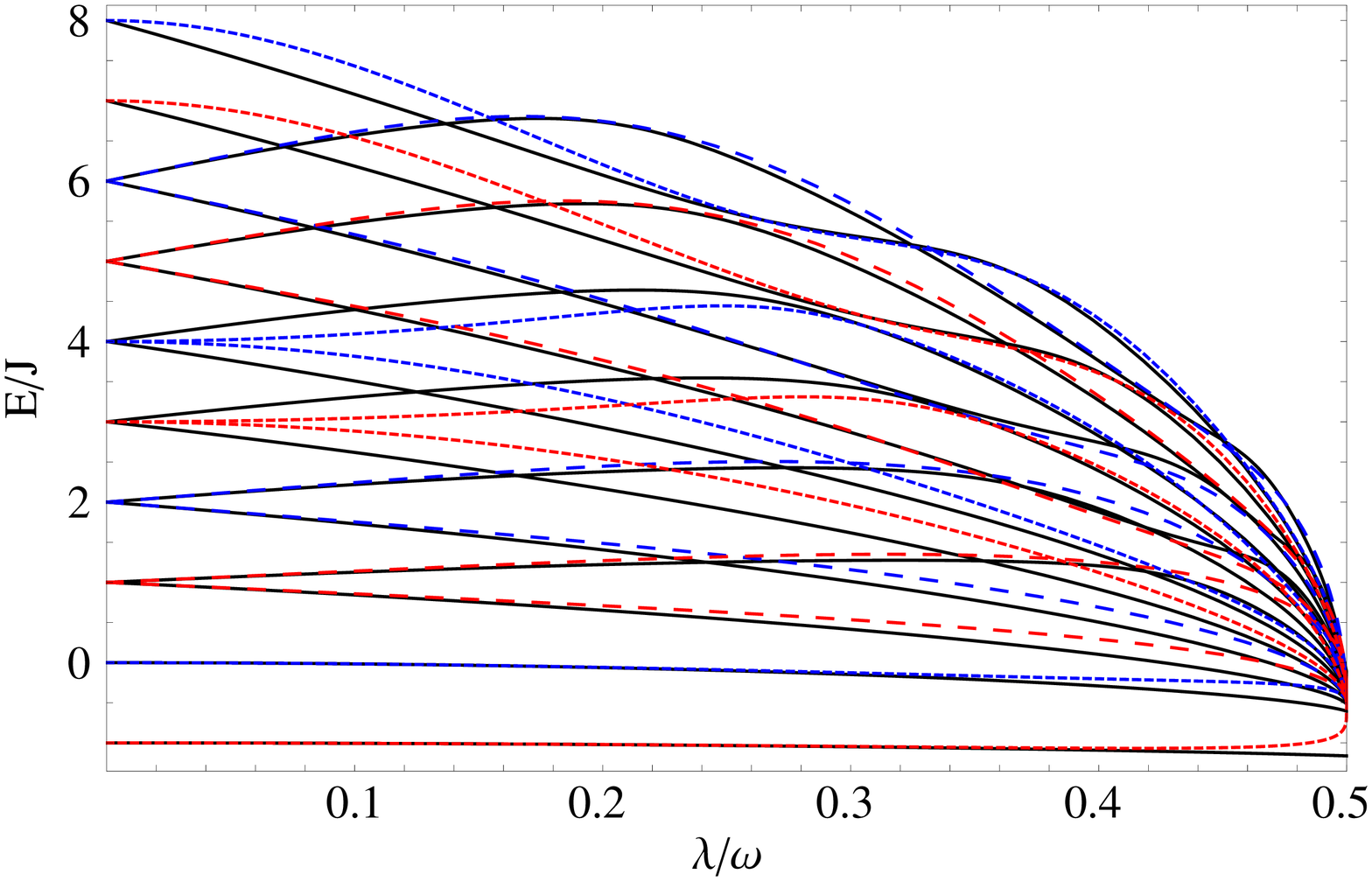}}
 \subfloat(b){\includegraphics[width=0.45\textwidth]{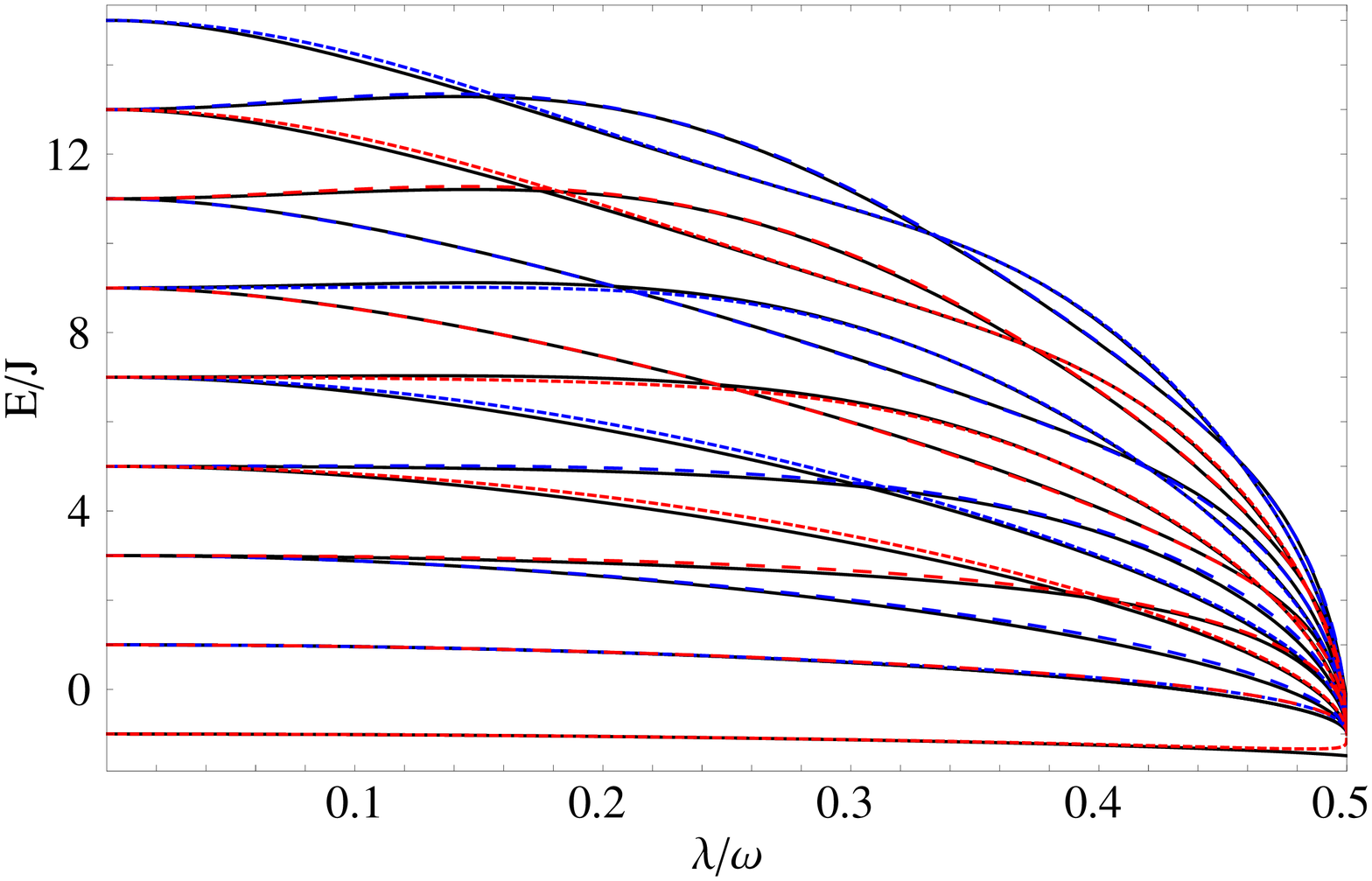}}
 \end{center}
\caption{(color online) Numerical energies (solid black) compared with the S-GRWA energies (dashed) (a) at the resonance case $ J = \omega = 1 $ and (b) at $ 2J = \omega=1$. The even (odd) manifold energies are red (blue); the respective positive (negative) parities on each manifold are in large (small) dashes. Although the numerical energies have the same parities and come from the same respective even/odd manifolds as the analytical approximation, all numerical results are in black for an easier visual comparison to the approximation. The numerical energies have been obtained by approximately diagonalizing the four matrices in Eq. (\ref{eq:fg2}) truncated to 100. \label{f3}}
\end{figure}

Diagonalizing these matrices gives two sets of symmetric GRWA energies for the two-photon Rabi Hamiltonian (with parity index $p=\pm$):
\begin{widetext}
\begin{eqnarray}
E_{N,p,\pm}^{\text{TP},even}&=&\left(4N+\frac{3}{2}\right)\tilde{\omega}-\frac{\omega}{2}+p\frac{J}{2}\sqrt{\tilde{\omega}/\omega}\left[P_{4N}\left(\tilde{\omega}/\omega\right)-P_{4N+2}\left(\tilde{\omega}/\omega\right)\right]\\
\nonumber
&&\pm\left(\left\{ \tilde{\omega}-p\frac{J}{2}\sqrt{\tilde{\omega}/\omega}\left[P_{4N}\left(\tilde{\omega}/\omega\right)+P_{4N+2}\left(\tilde{\omega}/\omega\right)\right]\right\} ^{2}+\frac{\tilde{\omega}J^{2}}{\left(4N+2\right)\left(4N+1\right)\omega}\left[P_{4N+1}^{1}\left(\tilde{\omega}/\omega\right)\right]^{2}\right)^{1/2}\\
E_{N,p,\pm}^{\text{TP},odd}&=&\left(4N+\frac{5}{2}\right)\tilde{\omega}-\frac{\omega}{2}+p\frac{J}{2}\sqrt{\tilde{\omega}/\omega}\left[P_{4N+1}\left(\tilde{\omega}/\omega\right)-P_{4N+3}\left(\tilde{\omega}/\omega\right)\right]\\
\nonumber
&&\pm\left(\left\{ \tilde{\omega}-p\frac{J}{2}\sqrt{\tilde{\omega}/\omega}\left[P_{4N+1}\left(\tilde{\omega}/\omega\right)+P_{4N+3}\left(\tilde{\omega}/\omega\right)\right]\right\} ^{2}+\frac{\tilde{\omega}J^{2}}{\left(4N+3\right)\left(4N+2\right)\omega}\left[P_{4N+2}^{1}\left(\tilde{\omega}/\omega\right)\right]^{2}\right)^{1/2}
\end{eqnarray}
\end{widetext}
These energies are plotted in Fig. \ref{f3} against the numerically obtained curves (black) for (a) $J=\omega=1$ and (b) $2J=\omega=1$. Since there are two orthogonal manifolds (even and odd) and a parity ($\pm$) for each, there are four different types of curves. Red (blue) denotes the even (odd) manifold and large (small) dashes are used for positive (negative) parity on each respective manifold. Although the numerical energies have the same parities and come from the same respective even/odd manifolds as the analytical approximation, all numerical results are in black for an easier visual comparison to the approximation. In the resonance case [Fig. \ref{f3} (a)], the respective even and odd manifold ground states are modeled well, with the ground state of the whole system being the ground state of the even manifold, $E_{0,-,-}^{\text{TP},even}$, for all values of the coupling. However, if one is to have an initial state consisting of strictly odd Fock states, then that state would evolve in a system whose ground state is that of the odd manifold, i.e., $E_{0,-,-}^{\text{TP},odd}$. Similar to the spin-boson case, the negative parity energies are not very accurate at small values of $\lambda$. However, the errors in energies for small coupling decrease with increasing energy level, just like for $H_1$. Another similarity to $H_1$ is that intersections occur between states for each respective manifold and between manifolds. The exact determination of the parity of each curve allows one to clarify the relative meaningfulness of intersections between them, building on previous efforts which have analytically determined these ``Juddian points'' \cite{tprh_bishop_exact, tprh_special_functions, *tprh_algebras_dolya}. Additionally, the form of the Hamiltonian in Eq. (\ref{eq:fg3}) allows one to see that $\widetilde{H}_{2}$ approaches $-\half \omega$ as $\lambda \rightarrow \half\omega$. In the off-resonance case of $2J = \omega = 1$ [Fig. \ref{f3} (b)], the results are significantly better and the different crossing pattern between parities and manifolds is reproduced by the analytical energies.

\section{Conclusion}

We have used a previously known unitary transformation to
diagonalize the single-mode spin-boson Hamiltonian in the
spin subspace and obtain symmetric versions of the regular
and generalized rotating wave approximations that respect
the symmetries of the Hamiltonian and include an equal
amount of rotating and counter-rotating contributions.
Additionally, we have devised an analogous unitary
transformation for the two-photon Rabi Hamiltonian and
obtained its respective symmetric generalized rotating wave
energies. The symmetric rotating wave approximation (S-RWA)
allows the short-time dynamics of the single-mode spin boson
and two-photon Rabi Hamiltonians to be analyzed in the ultra
and deep-strong coupling regimes with the relative numerical
simplicity of RWA-type approximations. The truncated S-GRWA
energies shown in this work and the two other proposed
higher-order truncations can be used to provide a highly
accurate quantitative picture of the details of the
dynamical behavior for both Hamiltonians. Both the S-RWA and
S-GRWA can be applied to model experiments where there is a
large coupling between the spin and bosonic systems and
where counter-rotating terms and parity are important
\cite{spin_boson_counter_rotating,
spin_boson_deep_strong_coupling,   grifoni2,
spin_boson_bloch_siegert,  spin_boson_zeno_physics,
spin_boson_zeno}.

The invariances with respect to the signs of $J$ and $\lambda$ present in the uncoupled Hamiltonians $H^{(0)}$ and $H_m^\pm$ ($m=1,2$), respectively, demonstrate that the concept of ``parity symmetry'' for the coupled systems $H_1$ and $H_2$ stems from the symmetries of these uncoupled Hamiltonians. Bosonic Hamiltonians $H_m^\pm$ with $m>2$ represent anharmonic oscillators and cannot be completely diagonalized by operators similar to $\mathcal{D}$ and $\mathcal{S}$ \cite{wagner_book}. As a result, the generalized treatment of the approximation in this work cannot be applied to the
ill-defined
\cite{multiphoton_ng_lui_pseudo-diagonalization} $m$-photon systems $H_m$ with $m>2$. Nevertheless, some of the anharmonic boson Hamiltonians $H_m^\pm$ do contain a definite parity symmetry: $\mathcal{R}$ is a parity operator for the odd anharmonic cases $m=3,5,7,...$ and $\mathcal{T}$ is a parity for $m=6,10,14,...\,$. The question of parity for the remaining cases $m=4,8,12,...$ is a subject of current investigation.


\begin{acknowledgments}
VVA thanks Hoda Hossein-Nejad for fruitful discussions. This work was supported by the Fulbright Canada Program
and by the U.S. Air Force Office of Scientific Research under contract number
FA9550-10-1-0260.

\end{acknowledgments}

\linespread{1}    
\bibliographystyle{apsrev4-1} 
\bibliography{C:/Users/Anil_Smith/Documents/VVA_Documents/RESEARCH/Papers/library} 


\end{document}